\documentclass[conference,10pt]{IEEEtran}
\usepackage{graphicx}
\usepackage[cmex10]{amsmath}
\usepackage[caption=false,font=footnotesize]{subfig}
\usepackage{color}
\usepackage{txfonts}
\usepackage{pgfplots}
\usepackage{booktabs,array}
\usepackage{multirow}
\usepackage{url}
\usepackage{enumerate}
\newcommand{\mus}{$\,\muup$s}
\renewcommand{\eqref}[1]{Eq.~(\ref{#1})}

\newcommand{\SC}{\mathcal{S}}
\newcommand{\Sa}{\mathcal{S}_1}
\newcommand{\Sb}{\mathcal{S}_2}
\newcommand{\N}{\mathcal{N}}

\newcommand{\Fa}{\mathcal{F}_1}
\newcommand{\Fb}{\mathcal{F}_2}

\DeclareMathOperator*{\argmin}{\arg min}

\graphicspath{{figures/}}
\newcommand{\subparagraph}{}
\usepackage{titlesec}
\titlespacing{\section}{0pt}{1.5ex plus 1.5ex minus 0.5ex}{0.7ex plus 1ex 
minus 0ex}
\begin{document}
    
\title{\hspace{-.15cm}RCFD: A Frequency--Based Channel Access 
Scheme\\ for 
Full--Duplex 
Wireless Networks}

\author{\IEEEauthorblockN{Michele Luvisotto$^\#$, Alireza Sadeghi$^{*}$, 
Farshad Lahouti$^{+}$, Stefano Vitturi$^\dag$, Michele Zorzi$^\#$}

	\IEEEauthorblockA{$^\#$Department of Information Engineering, University of 
	Padova, Italy}
	\IEEEauthorblockA{$^*$Electrical and Computer Engineering Department, 
	University of Minnesota, USA\\ }
	\IEEEauthorblockA{$^+$Electrical Engineering Department, California 
	Institute of Technology, USA\\ }
	\IEEEauthorblockA{$^\dag$CNR-IEIIT,	National Research Council of Italy, 
	Padova, Italy\\ }
	{\{luvisott, zorzi, vitturi\}@dei.unipd.it, sadeg012@umn.edu, 
	lahouti@caltech.edu}}

\maketitle


\begin{abstract}
Recently, several working implementations of in--band full--duplex wireless 
systems have been presented, where the same node can transmit and 
receive simultaneously in the same frequency band. 
The introduction of such a possibility at the physical layer could lead to 
improved performance but also poses several challenges at the MAC layer. In 
this paper, an innovative mechanism of channel 
contention in full--duplex OFDM wireless networks is proposed. 
This strategy is able to ensure efficient transmission scheduling with the 
result of avoiding collisions and effectively exploiting full--duplex 
opportunities. 
As a consequence, considerable performance improvements are observed with 
respect to standard and state--of--the--art MAC protocols for wireless 
networks, as highlighted by extensive simulations performed in ad hoc wireless 
networks with varying number of nodes. 
\end{abstract}

\section{Introduction}
\label{sec:introduction}
The currently employed channel access strategies in wireless networks provide 
good performance, namely high data rates and low latency, while ensuring 
acceptable fairness to all the users. However, they still suffer from some 
issues, such as the hidden and exposed terminal problems.
One of the most important limitations of wireless networks, with respect to 
wired solutions, is the so--called ``half--duplex constraint," i.e., the 
impossibility to transmit and receive in the same frequency band at the same 
time. 
The main challenge in obtaining full--duplex (FD) communication is the 
self--interference (SI) that affects the receive chain when transmission and 
reception occur simultaneously. 
The power of the signal emitted by a given node at its own receiver is much 
higher than that received from any other node, simply because the former comes 
from a much nearer source, with the result of completely destroying the signal 
of interest and prevent its successful decoding.

Recently, many research groups reported different implementations of 
self--interference cancellation schemes \cite{Choi2010,Duarte2010}, in order to 
eliminate the impact of the SI signal on the receiver and enable 
simultaneous transmission and reception in the same frequency band. 
These findings opened the way for a new branch of research, 
concerned with redesigning the higher layers of wireless networks in order to 
fully exploit these new FD capabilities.
In particular, the definition of MAC layer strategies able to take 
advantage of such new features is a topic of great interest. Although numerous 
schemes have already been proposed, for a variety of network architectures 
\cite{Duarte2014,Sahai2011,Kim2013}, all of them present 
specific benefits and drawbacks and no proposal has emerged as the leading 
one for the definition of a future FD MAC layer protocol. 
It has to be remarked 
that the possibility for a node to receive and transmit at the same time, 
although allowing a potential doubling of the network capacity, increases the 
nodes exposition to interference and considerably complicates the scheduling of 
transmissions. 
The study of channel access schemes capable of efficiently exploiting the FD 
capabilities and producing significant performance gains compared to current 
wireless systems is, consequently, a very important research topic.

Another direction of research that aims at overcoming the limitations of
standard, time--domain based, distributed channel access schemes for wireless 
networks relies on the idea of moving the entire channel contention 
procedure to the frequency domain \cite{Sen2010}. The 
proposed solution exploits the availability of multiple subcarriers (SCs), 
derived from the common adoption of Orthogonal Frequency--Division Multiplexing 
(OFDM) modulation in modern wireless networks, and lets the nodes contend for 
the channel by randomly selecting one of these SCs. This innovative approach 
allows to solve contention in a short amount of time compared to traditional 
time--domain schemes, such as the Carrier Sense Multiple Access with Collision 
Avoidance (CSMA/CA)  adopted in IEEE 802.11 networks. However, it
faces some limitations and, in particular, appears to be unsuitable for the 
realistic case of a network with multiple contention domains. With the term 
collision (or contention) domain of a node, we refer to the set of nodes in the 
network that are within its coverage area. In a network with a single collision 
domain, the coverage area of each node includes all the other nodes in the 
network, whereas in the case of multiple collision domains, nodes have 
different coverage areas.

In this paper, we revisit the frequency--domain contention approach and propose 
a new channel access algorithm able to efficiently schedule 
transmissions in a context where nodes are equipped with FD capabilities.  
The scheme works in general ad hoc wireless networks, is fully distributed, 
preserves the desired randomness thus ensuring fairness among different nodes, 
and is not affected by the presence of several contention domains. 
Indeed, multiple contention rounds in the frequency domain, each corresponding 
to an OFDM symbol, are used to advertise the transmission intentions of the 
nodes and to select the pair of nodes that will actually perform a data 
exchange 
within a single collision domain. Since this approach resembles the Request To 
Send/Clear To Send (RTS/CTS) scheme designed for IEEE 802.11 networks 
\cite{ieee80211std}, we called the new MAC layer scheme as RTS/CTS in the
Frequency Domain (RCFD). Such an approach eliminates some of the 
problems that traditionally affect wireless networks, for example the hidden 
terminal (HT) issue.

The rest of the paper is organized as follows. Section~\ref{sec:related} 
presents the state--of--the--art on the topics discussed in 
this paper and introduces some general concepts. The structure of the proposed 
RCFD MAC protocol is given in Section~\ref{sec:protocol}, together with some 
examples of operations. The validation of this strategy against other MAC 
layer 
schemes is provided in Section~\ref{sec:simulation} through
simulation results. Finally, Section~\ref{sec:conclusions} 
concludes the paper.

\section{Background and Related Works}
\label{sec:related}

In this Section, we discuss some preliminary concepts and refer to several 
contributions that appeared in the scientific literature concerned with both 
full--duplex wireless communication and frequency--based channel access, whose 
combination is the focus of this paper.

\subsection{Full--duplex Wireless}
\label{subsec:related_fd}

In the last decade, several researchers
have worked on different techniques to cope with SI and enable FD wireless, 
until 2010 when some research groups independently presented the first 
functioning prototypes \cite{Choi2010,Duarte2010,Radunovic2010}.

An exhaustive summary about in--band FD wireless can be found in 
\cite{Sabharwal2013}, \cite{Kim2015}.  These works present in detail 
the various available PHY layer methods to suppress SI, as well as the 
main challenges and research opportunities in this field, and specifically 
discuss the research issues at the MAC layer. 

Full--duplex MAC protocols have been designed for both infrastructure and 
ad hoc wireless networks. In the infrastructure configuration, some 
schemes have been developed for the case of asymmetric traffic, such as 
\cite{Jain2011}, \cite{Sahai2011}, that do not consider any interference 
between nodes, or \cite{Kim2013}, that takes this issue into account and 
proposes a centralized scheduler.
More strategies are available for ad hoc networks, e.g., 
\cite{Singh2011}, which proposes a distributed scheduling 
protocol aimed at enhancing fairness, or \cite{Duarte2014,Cheng2013a}, 
which make use of RTS/CTS packets. 
Other works propose solutions to enhance the end--to--end performance of 
multi--hop 
FD networks, such as \cite{Miura2012}, where the use of directional antennas is 
addressed, \cite{Sadeghi2014}, where frequency reuse to enhance outage 
probability is investigated, and \cite{Tamaki2013}, that proposes asynchronous 
channel access. 

\subsection{Frequency--based Channel Access}

The concept of moving channel contention to the frequency domain was first 
introduced in \cite{Sen2010}. 
The authors proposed a simple scheme for OFDM--based systems: 
the channel access procedure starts with a contention round in which each 
contending node randomly selects an SC among those available and transmits a 
symbol only on that portion of the spectrum, while listening to the whole band. 
If the chosen SC is the first one\footnote{In this paper we refer to the 
``first'' subcarrier as the one with the lowest frequency.} that carries a 
signal, the node grabs the 
channel and is allowed to transmit, whereas the other nodes (that have chosen 
other SCs) remain silent. 
A similar strategy was discussed in \cite{Zhang2012}, where the set of 
available SCs is divided in two subsets, one used for random 
contention and the other for node identification. 

\section{The RCFD Full--duplex MAC Protocol}
\label{sec:protocol}

The proposed RCFD algorithm is a frequency--based channel access scheme, 
in which not only the medium contention, but also transmission identification 
and selection are performed in the frequency domain, similar to the RTS/CTS 
procedure usually performed in the time domain. 

\subsection{Preliminaries}
\label{subsec:assumptions}

The proposed protocol relies on some assumptions that ensure its correct 
behavior. 

RCFD is designed for a general ad hoc wireless network, formed by 
independent nodes that have the same priority. It is a distributed scheme where 
no central coordination is required.  

Each node is assumed to have FD capabilities. In this work, we do not consider 
any residual effect of SI, since we assume that an advanced FD terminal is 
adopted, able to reduce the SI level to the noise floor for the frequency bands 
of interest \cite{Bharadia2013}.
We also suppose, for the sake of clarity and only in this first description, 
that all the nodes try to access the channel simultaneously and a non--empty 
subset of them has data to send. 

The communication channel is assumed ideal (no external interference, fading or 
path loss), so that each node can hear every other node within its coverage 
range. There can be multiple collision domains, i.e., the communication range 
of a node may not include all the nodes in the network.

The most important assumption is that a unique association between 
each node and two OFDM subcarriers is initially established at network setup 
and maintained fixed throughout all operations. More specifically, defining 
$\SC=\{s_1,\dots,s_S\}$ as the set of available SCs, we split it in two 
non--overlapping parts $\Sa$ and $\Sb$. Taking
$\N=\{n_1,\dots,n_N\}$ as the set of network nodes, a unique 
mapping is defined by the two functions
\begin{equation}
\Fa: \N \to \Sa,\quad\Fb : \N \to \Sb
\end{equation}
that link each node with two subcarriers in $\Sa$ and $\Sb$ uniquely associated 
to it. Specifically, we must have $\Fa(n_i)\neq\Fa(n_j)$ and 
$\Fb(n_i)\neq\Fb(n_j)$ for any $i\neq j$.
As an example, the simplest implementation of such a mapping is obtained if we 
take $\Sa=\{s_1,\dots,s_{S/2}\}$, $\Sb=\{s_{S/2+1},\dots,s_{S}\}$ and define 
$\Fa(n_i)=s_i,\;\Fb(n_i)=s_{i+S/2},\;i=1,\dots,N$.

The partition of the available SCs into two subsets is required since, during 
the contention procedure, every node should advertise two different 
information, namely its ID and the ID of its receiver, as will be better 
detailed in the rest of this Section.
It should be noted that the assumed subcarrier mapping actually imposes a 
constraint on the number of nodes in the network. Indeed, since each node must 
be uniquely associated with two OFDM SCs, the total number of nodes has to be 
less than or equal to $S/2$. The impact of this assumption and how it could be 
relaxed are discussed in Section~\ref{subsec:discussion}.

\subsection{Channel Contention Scheme}
\label{subsec:channel_access}

The channel access procedure is composed of three contention rounds in the 
frequency domain. The first round starts after each node has sensed the channel 
and found it idle for a certain period of time $T_{scan}$. Each round involves 
the transmission of an OFDM symbol and its duration is set to 
$T_{round}=T_{sym}+2T_p$ to accommodate for signal propagation, which takes a 
time $T_p$ \cite{Sen2011}. Therefore, the channel access procedure takes a 
fixed time $T_{acc}=T_{scan}+3\cdot T_{round}$. As an example, in IEEE 
802.11g networks standard values are $T_{scan}=$ 28~\mus\ (a Distributed 
Inter--Frame Space interval), $T_{sym}=$ 4~\mus, $T_p=$ 1~\mus, thus obtaining 
$T_{acc}=$ 46~\mus.

In the following, we outline the steps performed by every node in each 
contention round.

\subsubsection{First round - randomized contention}

Every node that has data to send and has found the channel idle for a 
$T_{scan}$ period randomly selects an SC from the whole set $\SC$ and transmits 
a symbol only on that SC, while listening to the whole band. It is worth noting 
that also simple HD nodes can listen to the whole channel while 
transmitting on a single SC, but the listening procedure is affected by SC 
leakage, that makes it difficult to detect signals received on SCs adjacent to 
the one selected for transmission \cite{Sen2011}. Conversely, an FD terminal 
can 
be much more precise in that it can cancel the effect of a transmission
limited to an SC on the receive path, thus eliminating the SC leakage.
We denote with $\bar{s}_i\in\SC$ the SC chosen by node $n_i,\;i=1,\dots,N$, 
where $\bar{s}_i=0$ if 
node $n_i$ does not have data to send. We also indicate with $\SC_i^1$ the set 
of SCs that actually carried a symbol during the first contention round, as 
perceived by node $n_i$. 

Node $n_i$ is defined as \textit{primary transmitter} (PT) if and only if the 
following condition holds
\begin{equation}
\bar{s}_i=\min\left[\SC_i^1\right]
\end{equation}
i.e., the first SC carrying data is the one chosen by the node itself. It has 
to be remarked that, in the realistic scenario of multiple collision domains, 
several nodes in the network can be selected as PTs. Moreover, it can also 
happen that multiple nodes in the same collision domain pick the same SC. In 
this case, they are both selected as PTs and the following rounds will be used 
to select the actual transmitter.

\subsubsection{Second round - transmission advertisement (RTS)}

In the second round, nodes selected as PTs in the first round will advertise 
their transmission intentions. This is the so--called Request--To--Send (RTS) 
part of the algorithm.  

A PT node $n_i$ that has data for node $n_j$ transmits a symbol on two 
SCs, namely $s_a=\Fa\left(n_i\right)$ and 
$s_b=\Fb(n_j)$. In this way, $n_i$ informs its neighbors that it is a PT and 
has a packet for $n_j$.

All the nodes in the network, including the other PTs, listen to the whole band 
during the second round. We denote as $\SC_{h,1}^2\subseteq\Sa$ and 
$\SC_{h,2}^2\subseteq\Sb$ the sets of SCs that carried a symbol during the 
second contention round, as perceived by a generic node $n_h$. 

Node $n_h$ is defined as \textit{RTS receiver} (RR) if and only if the 
following condition holds
\begin{equation}
\label{eq:rr}
\Fb(n_h)\in\SC_{h,2}^2
\end{equation}
i.e., another node, who has been selected as a PT, informed that it has a 
packet for $n_h$. There can be multiple RRs in the network, but a node cannot 
be both PT and RR at the same time, since, in order to be a PT, all the nodes 
within its coverage range must not be PTs, therefore they could not send an RTS.

\begin{figure*}[!t]
\begin{center}
\scalebox{0.7}{
\begin{tikzpicture}
	\node at (0,-3) {Node $n_1$};
	\draw[draw=black,very thick] (1,-3)--(7,-3);
	\draw (1.5,-3) -- (1.5,-3.2);
	\node at (1.5,-3.4) {$s_1$};
	\draw (2.5,-3) -- (2.5,-3.2);
	\node at (2.5,-3.4) {$s_2$};
	\draw (3.5,-3) -- (3.5,-3.2);
	\node at (3.5,-3.4) {$s_3$};
	\draw (4.5,-3) -- (4.5,-3.2);
	\node at (4.5,-3.4) {$s_4$};
	\draw (5.5,-3) -- (5.5,-3.2);
	\node at (5.5,-3.4) {$s_5$};
	\draw (6.5,-3) -- (6.5,-3.2);
	\node at (6.5,-3.4) {$s_6$};	
	\draw[-latex, ultra thick,draw=green,fill=green] (1.5,-3) -- (1.5,-2);
	\node at (0,-4.5) {Node $n_2$};
	\draw[draw=black,very thick] (1,-4.5)--(7,-4.5);
	\draw (1.5,-4.5) -- (1.5,-4.7);
	\node at (1.5,-4.9) {$s_1$};
	\draw (2.5,-4.5) -- (2.5,-4.7);
	\node at (2.5,-4.9) {$s_2$};
	\draw (3.5,-4.5) -- (3.5,-4.7);
	\node at (3.5,-4.9) {$s_3$};
	\draw (4.5,-4.5) -- (4.5,-4.7);
	\node at (4.5,-4.9) {$s_4$};
	\draw (5.5,-4.5) -- (5.5,-4.7);
	\node at (5.5,-4.9) {$s_5$};
	\draw (6.5,-4.5) -- (6.5,-4.7);
	\node at (6.5,-4.9) {$s_6$};	
	\draw[-latex, ultra thick,draw=red,fill=red] (1.5,-4.5) -- (1.5,-4);
	\draw[-latex, ultra thick,draw=red,fill=red] (3.5,-4.5) -- (3.5,-4);
	\node at (0,-6) {Node $n_3$};
	\draw[draw=black,very thick] (1,-6)--(7,-6);
	\draw (1.5,-6) -- (1.5,-6.2);
	\node at (1.5,-6.4) {$s_1$};
	\draw (2.5,-6) -- (2.5,-6.2);
	\node at (2.5,-6.4) {$s_2$};
	\draw (3.5,-6) -- (3.5,-6.2);
	\node at (3.5,-6.4) {$s_3$};
	\draw (4.5,-6) -- (4.5,-6.2);
	\node at (4.5,-6.4) {$s_4$};
	\draw (5.5,-6) -- (5.5,-6.2);
	\node at (5.5,-6.4) {$s_5$};
	\draw (6.5,-6) -- (6.5,-6.2);
	\node at (6.5,-6.4) {$s_6$};
	\draw[-latex, ultra thick,draw=green,fill=green] (3.5,-6) -- (3.5,-5);
	\node at (4,-7) {First round};
	\draw[draw=black,very thick] (8,-3)--(14,-3);
	\draw (8.5,-3) -- (8.5,-3.2);
	\node at (8.5,-3.4) {$s_1$};
	\draw (9.5,-3) -- (9.5,-3.2);
	\node at (9.5,-3.4) {$s_2$};
	\draw (10.5,-3) -- (10.5,-3.2);
	\node at (10.5,-3.4) {$s_3$};
	\draw (11.5,-3) -- (11.5,-3.2);
	\node at (11.5,-3.4) {$s_4$};
	\draw (12.5,-3) -- (12.5,-3.2);
	\node at (12.5,-3.4) {$s_5$};
	\draw (13.5,-3) -- (13.5,-3.2);
	\node at (13.5,-3.4) {$s_6$};	
	\draw[-latex, ultra thick,draw=green,fill=green] (8.5,-3) -- (8.5,-2);
	\draw[-latex, ultra thick,draw=green,fill=green] (12.5,-3) -- (12.5,-2);
	\draw[draw=black,very thick] (8,-4.5)--(14,-4.5);
	\draw (8.5,-4.5) -- (8.5,-4.7);
	\node at (8.5,-4.9) {$s_1$};
	\draw (9.5,-4.5) -- (9.5,-4.7);
	\node at (9.5,-4.9) {$s_2$};
	\draw (10.5,-4.5) -- (10.5,-4.7);
	\node at (10.5,-4.9) {$s_3$};
	\draw (11.5,-4.5) -- (11.5,-4.7);
	\node at (11.5,-4.9) {$s_4$};
	\draw (12.5,-4.5) -- (12.5,-4.7);
	\node at (12.5,-4.9) {$s_5$};
	\draw (13.5,-4.5) -- (13.5,-4.7);
	\node at (13.5,-4.9) {$s_6$};
	\draw[-latex, ultra thick,draw=red,fill=red] (8.5,-4.5) -- (8.5,-4);
	\draw[-latex, ultra thick,draw=red,fill=red] (10.5,-4.5) -- (10.5,-4);	
	\draw[-latex, ultra thick,draw=red,fill=red] (12.5,-4.5) -- (12.5,-4);	
	\draw[draw=black,very thick] (8,-6)--(14,-6);
	\draw (8.5,-6) -- (8.5,-6.2);
	\node at (8.5,-6.4) {$s_1$};
	\draw (9.5,-6) -- (9.5,-6.2);
	\node at (9.5,-6.4) {$s_2$};
	\draw (10.5,-6) -- (10.5,-6.2);
	\node at (10.5,-6.4) {$s_3$};
	\draw (11.5,-6) -- (11.5,-6.2);
	\node at (11.5,-6.4) {$s_4$};
	\draw (12.5,-6) -- (12.5,-6.2);
	\node at (12.5,-6.4) {$s_5$};
	\draw (13.5,-6) -- (13.5,-6.2);
	\node at (13.5,-6.4) {$s_6$};
	\draw[-latex, ultra thick,draw=green,fill=green] (10.5,-6) -- (10.5,-5);
	\draw[-latex, ultra thick,draw=green,fill=green] (12.5,-6) -- (12.5,-5);
	\node at (11,-7) {Second round};
	\draw[draw=black,very thick] (15,-3)--(21,-3);
	\draw (15.5,-3) -- (15.5,-3.2);
	\node at (15.5,-3.4) {$s_1$};
	\draw (16.5,-3) -- (16.5,-3.2);
	\node at (16.5,-3.4) {$s_2$};
	\draw (17.5,-3) -- (17.5,-3.2);
	\node at (17.5,-3.4) {$s_3$};
	\draw (18.5,-3) -- (18.5,-3.2);
	\node at (18.5,-3.4) {$s_4$};
	\draw (19.5,-3) -- (19.5,-3.2);
	\node at (19.5,-3.4) {$s_5$};
	\draw (20.5,-3) -- (20.5,-3.2);
	\node at (20.5,-3.4) {$s_6$};
	\draw[-latex, ultra thick,draw=red,fill=red] (16.5,-3) -- (16.5,-2.5);	
	\draw[-latex, ultra thick,draw=red,fill=red] (18.5,-3) -- (18.5,-2.5);	
	\draw[draw=black,very thick] (15,-4.5)--(21,-4.5);
	\draw (15.5,-4.5) -- (15.5,-4.7);
	\node at (15.5,-4.9) {$s_1$};
	\draw (16.5,-4.5) -- (16.5,-4.7);
	\node at (16.5,-4.9) {$s_2$};
	\draw (17.5,-4.5) -- (17.5,-4.7);
	\node at (17.5,-4.9) {$s_3$};
	\draw (18.5,-4.5) -- (18.5,-4.7);
	\node at (18.5,-4.9) {$s_4$};
	\draw (19.5,-4.5) -- (19.5,-4.7);
	\node at (19.5,-4.9) {$s_5$};
	\draw (20.5,-4.5) -- (20.5,-4.7);
	\node at (20.5,-4.9) {$s_6$};	
	\draw[-latex, ultra thick,draw=green,fill=green] (16.5,-4.5) -- 
	(16.5,-3.5);	
	\draw[-latex, ultra thick,draw=green,fill=green] (18.5,-4.5) -- (18.5,-3.5);
	\draw[draw=black,very thick] (15,-6)--(21,-6);
	\draw (15.5,-6) -- (15.5,-6.2);
	\node at (15.5,-6.4) {$s_1$};
	\draw (16.5,-6) -- (16.5,-6.2);
	\node at (16.5,-6.4) {$s_2$};
	\draw (17.5,-6) -- (17.5,-6.2);
	\node at (17.5,-6.4) {$s_3$};
	\draw (18.5,-6) -- (18.5,-6.2);
	\node at (18.5,-6.4) {$s_4$};
	\draw (19.5,-6) -- (19.5,-6.2);
	\node at (19.5,-6.4) {$s_5$};
	\draw (20.5,-6) -- (20.5,-6.2);
	\node at (20.5,-6.4) {$s_6$};
	\draw[-latex, ultra thick,draw=red,fill=red] (16.5,-6) -- (16.5,-5.5);	
	\draw[-latex, ultra thick,draw=red,fill=red] (18.5,-6) -- (18.5,-5.5);	
	\node at (18,-7) {Third round};
	\node at (23,-3) {\textcolor{green}{TX subcarriers}};
	\node at (23.25,-3.75) {\textcolor{red}{Heard subcarriers}};	
\end{tikzpicture}}
\end{center}
\caption{Outcomes of contention rounds for example Scenario 1.}
\label{fig:sc_scenario1}
\end{figure*}
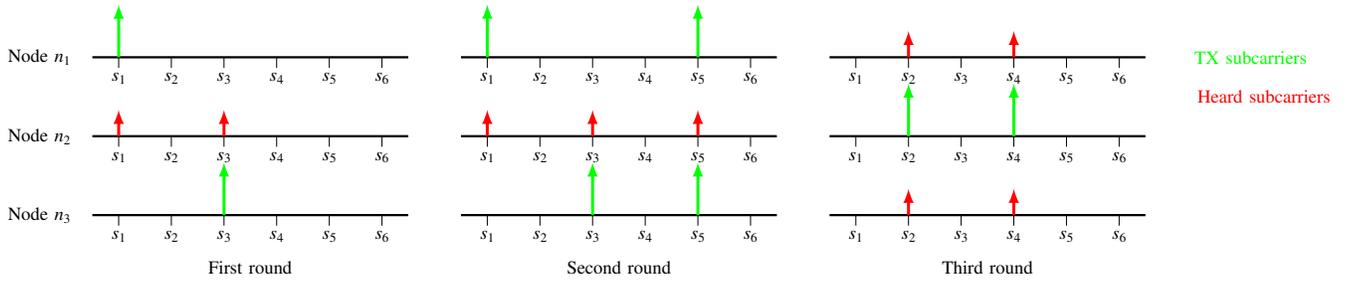
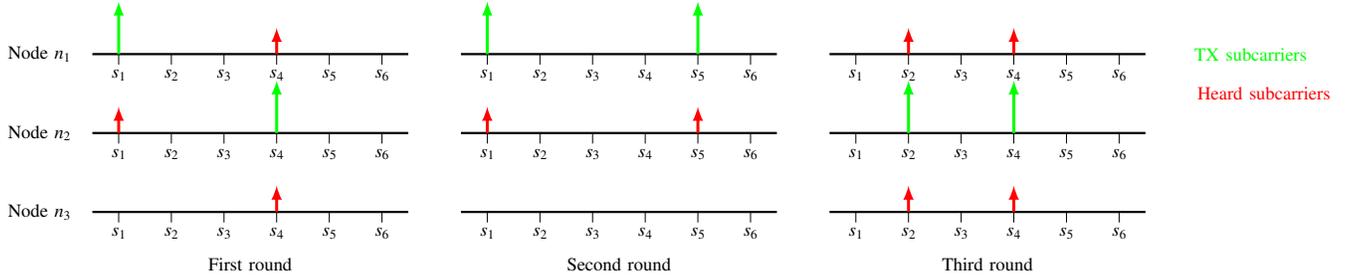
\begin{figure*}[!t]
\begin{center}
\scalebox{0.7}{
\begin{tikzpicture}
	\node at (0,-3) {Node $n_1$};
	\draw[draw=black,very thick] (1,-3)--(7,-3);
	\draw (1.5,-3) -- (1.5,-3.2);
	\node at (1.5,-3.4) {$s_1$};
	\draw (2.5,-3) -- (2.5,-3.2);
	\node at (2.5,-3.4) {$s_2$};
	\draw (3.5,-3) -- (3.5,-3.2);
	\node at (3.5,-3.4) {$s_3$};
	\draw (4.5,-3) -- (4.5,-3.2);
	\node at (4.5,-3.4) {$s_4$};
	\draw (5.5,-3) -- (5.5,-3.2);
	\node at (5.5,-3.4) {$s_5$};
	\draw (6.5,-3) -- (6.5,-3.2);
	\node at (6.5,-3.4) {$s_6$};	
	\draw[-latex, ultra thick,draw=green,fill=green] (1.5,-3) -- (1.5,-2);
	\draw[-latex, ultra thick,draw=red,fill=red] (4.5,-3) -- (4.5,-2.5);
	\node at (0,-4.5) {Node $n_2$};
	\draw[draw=black,very thick] (1,-4.5)--(7,-4.5);
	\draw (1.5,-4.5) -- (1.5,-4.7);
	\node at (1.5,-4.9) {$s_1$};
	\draw (2.5,-4.5) -- (2.5,-4.7);
	\node at (2.5,-4.9) {$s_2$};
	\draw (3.5,-4.5) -- (3.5,-4.7);
	\node at (3.5,-4.9) {$s_3$};
	\draw (4.5,-4.5) -- (4.5,-4.7);
	\node at (4.5,-4.9) {$s_4$};
	\draw (5.5,-4.5) -- (5.5,-4.7);
	\node at (5.5,-4.9) {$s_5$};
	\draw (6.5,-4.5) -- (6.5,-4.7);
	\node at (6.5,-4.9) {$s_6$};	
	\draw[-latex, ultra thick,draw=red,fill=red] (1.5,-4.5) -- (1.5,-4);
	\draw[-latex, ultra thick,draw=green,fill=green] (4.5,-4.5) -- (4.5,-3.5);
	\node at (0,-6) {Node $n_3$};
	\draw[draw=black,very thick] (1,-6)--(7,-6);
	\draw (1.5,-6) -- (1.5,-6.2);
	\node at (1.5,-6.4) {$s_1$};
	\draw (2.5,-6) -- (2.5,-6.2);
	\node at (2.5,-6.4) {$s_2$};
	\draw (3.5,-6) -- (3.5,-6.2);
	\node at (3.5,-6.4) {$s_3$};
	\draw (4.5,-6) -- (4.5,-6.2);
	\node at (4.5,-6.4) {$s_4$};
	\draw (5.5,-6) -- (5.5,-6.2);
	\node at (5.5,-6.4) {$s_5$};
	\draw (6.5,-6) -- (6.5,-6.2);
	\node at (6.5,-6.4) {$s_6$};
	\draw[-latex, ultra thick,draw=red,fill=red] (4.5,-6) -- (4.5,-5.5);
	\node at (4,-7) {First round};
	\draw[draw=black,very thick] (8,-3)--(14,-3);
	\draw (8.5,-3) -- (8.5,-3.2);
	\node at (8.5,-3.4) {$s_1$};
	\draw (9.5,-3) -- (9.5,-3.2);
	\node at (9.5,-3.4) {$s_2$};
	\draw (10.5,-3) -- (10.5,-3.2);
	\node at (10.5,-3.4) {$s_3$};
	\draw (11.5,-3) -- (11.5,-3.2);
	\node at (11.5,-3.4) {$s_4$};
	\draw (12.5,-3) -- (12.5,-3.2);
	\node at (12.5,-3.4) {$s_5$};
	\draw (13.5,-3) -- (13.5,-3.2);
	\node at (13.5,-3.4) {$s_6$};	
	\draw[-latex, ultra thick,draw=green,fill=green] (8.5,-3) -- (8.5,-2);
	\draw[-latex, ultra thick,draw=green,fill=green] (12.5,-3) -- (12.5,-2);
	\draw[draw=black,very thick] (8,-4.5)--(14,-4.5);
	\draw (8.5,-4.5) -- (8.5,-4.7);
	\node at (8.5,-4.9) {$s_1$};
	\draw (9.5,-4.5) -- (9.5,-4.7);
	\node at (9.5,-4.9) {$s_2$};
	\draw (10.5,-4.5) -- (10.5,-4.7);
	\node at (10.5,-4.9) {$s_3$};
	\draw (11.5,-4.5) -- (11.5,-4.7);
	\node at (11.5,-4.9) {$s_4$};
	\draw (12.5,-4.5) -- (12.5,-4.7);
	\node at (12.5,-4.9) {$s_5$};
	\draw (13.5,-4.5) -- (13.5,-4.7);
	\node at (13.5,-4.9) {$s_6$};
	\draw[-latex, ultra thick,draw=red,fill=red] (8.5,-4.5) -- (8.5,-4);
	\draw[-latex, ultra thick,draw=red,fill=red] (12.5,-4.5) -- (12.5,-4);	
	\draw[draw=black,very thick] (8,-6)--(14,-6);
	\draw (8.5,-6) -- (8.5,-6.2);
	\node at (8.5,-6.4) {$s_1$};
	\draw (9.5,-6) -- (9.5,-6.2);
	\node at (9.5,-6.4) {$s_2$};
	\draw (10.5,-6) -- (10.5,-6.2);
	\node at (10.5,-6.4) {$s_3$};
	\draw (11.5,-6) -- (11.5,-6.2);
	\node at (11.5,-6.4) {$s_4$};
	\draw (12.5,-6) -- (12.5,-6.2);
	\node at (12.5,-6.4) {$s_5$};
	\draw (13.5,-6) -- (13.5,-6.2);
	\node at (13.5,-6.4) {$s_6$};
	\node at (11,-7) {Second round};
	\draw[draw=black,very thick] (15,-3)--(21,-3);
	\draw (15.5,-3) -- (15.5,-3.2);
	\node at (15.5,-3.4) {$s_1$};
	\draw (16.5,-3) -- (16.5,-3.2);
	\node at (16.5,-3.4) {$s_2$};
	\draw (17.5,-3) -- (17.5,-3.2);
	\node at (17.5,-3.4) {$s_3$};
	\draw (18.5,-3) -- (18.5,-3.2);
	\node at (18.5,-3.4) {$s_4$};
	\draw (19.5,-3) -- (19.5,-3.2);
	\node at (19.5,-3.4) {$s_5$};
	\draw (20.5,-3) -- (20.5,-3.2);
	\node at (20.5,-3.4) {$s_6$};
	\draw[-latex, ultra thick,draw=red,fill=red] (16.5,-3) -- (16.5,-2.5);	
	\draw[-latex, ultra thick,draw=red,fill=red] (18.5,-3) -- (18.5,-2.5);	
	\draw[draw=black,very thick] (15,-4.5)--(21,-4.5);
	\draw (15.5,-4.5) -- (15.5,-4.7);
	\node at (15.5,-4.9) {$s_1$};
	\draw (16.5,-4.5) -- (16.5,-4.7);
	\node at (16.5,-4.9) {$s_2$};
	\draw (17.5,-4.5) -- (17.5,-4.7);
	\node at (17.5,-4.9) {$s_3$};
	\draw (18.5,-4.5) -- (18.5,-4.7);
	\node at (18.5,-4.9) {$s_4$};
	\draw (19.5,-4.5) -- (19.5,-4.7);
	\node at (19.5,-4.9) {$s_5$};
	\draw (20.5,-4.5) -- (20.5,-4.7);
	\node at (20.5,-4.9) {$s_6$};	
	\draw[-latex, ultra thick,draw=green,fill=green] (16.5,-4.5) -- 
	(16.5,-3.5);	
	\draw[-latex, ultra thick,draw=green,fill=green] (18.5,-4.5) -- (18.5,-3.5);
	\draw[draw=black,very thick] (15,-6)--(21,-6);
	\draw (15.5,-6) -- (15.5,-6.2);
	\node at (15.5,-6.4) {$s_1$};
	\draw (16.5,-6) -- (16.5,-6.2);
	\node at (16.5,-6.4) {$s_2$};
	\draw (17.5,-6) -- (17.5,-6.2);
	\node at (17.5,-6.4) {$s_3$};
	\draw (18.5,-6) -- (18.5,-6.2);
	\node at (18.5,-6.4) {$s_4$};
	\draw (19.5,-6) -- (19.5,-6.2);
	\node at (19.5,-6.4) {$s_5$};
	\draw (20.5,-6) -- (20.5,-6.2);
	\node at (20.5,-6.4) {$s_6$};
	\draw[-latex, ultra thick,draw=red,fill=red] (16.5,-6) -- (16.5,-5.5);	
	\draw[-latex, ultra thick,draw=red,fill=red] (18.5,-6) -- (18.5,-5.5);	
	\node at (18,-7) {Third round};
	\node at (23,-3) {\textcolor{green}{TX subcarriers}};
	\node at (23.25,-3.75) {\textcolor{red}{Heard subcarriers}};	
\end{tikzpicture}}
\end{center}
\caption{Outcomes of contention rounds for example Scenario 2.}
\label{fig:sc_scenario2}
\end{figure*}

\subsubsection{Third round - transmission authorization (CTS)}

In the third round, nodes selected as RR in the second round will grant or 
refuse the permission to transmit. This is the so--called Clear--To--Send (CTS) 
part of the algorithm. 

An RR node $n_h$ will send the CTS to node
\begin{equation}
n_l = 
\argmin\limits_{n_i\in\mathcal{N}}\left[\Fa(n_i)\;:\;\Fa(n_i)\in\SC_{h,1}^2\right]
\end{equation}
i.e., among the nodes that have sent an RTS to $n_h$, it selects the one with 
the first associated SC in $\Sa$. Node $n_h$ then 
transmits a symbol on two SCs, namely $s_c=\Fa\left(n_h\right)$ and 
$s_d=\Fb(n_l)$. In this way, $n_h$ informs $n_l$ 
that its transmission is authorized. All the nodes in the network, including 
the RRs, listen to the whole band during the third contention round and we 
denote as $\SC_{i,1}^3\subseteq\Sa$ and $\SC_{i,2}^3\subseteq\Sb$ the sets of 
SCs that 
carried a symbol during the third contention round, as perceived by a generic 
node $n_i$.

At the end of the third round, each node that has data to send needs to decide 
whether to transmit or not, according to the information gathered in the three 
rounds. 
Here it is worth stressing that not only the nodes selected as PTs during the 
first round may be granted access to the channels, but also an RR can transmit, 
provided that some conditions are verified. This possibility is the key to 
enable FD transmission: a node that has a packet for another node from which it 
has received an RTS can send it together with the main transmission.

Specifically, assuming a generic node $n_i$ which has a packet for node $n_j$, 
three cases can be distinguished. If $n_i$ is a PT, it transmits if and only 
if both these conditions are verified
\begin{equation}
\label{eq:transm_pt}
\Fa(n_j)\in\SC_{i,1}^3,\quad
\SC_{i,2}^3=\left\{\Fb(n_i)\right\}
\end{equation}
i.e., the intended receiver (node $n_j$) has sent a CTS and this is the only 
CTS heard within the contention domain of node $n_i$. Else, if $n_i$ is an RR, 
it transmits if and only if both these conditions are verified
\begin{equation}
\label{eq:transm_rr}
\SC_{i,1}^2=\left\{\Fa(n_j)\right\},\quad
\SC_{i,1}^3=\left\{\Fa(n_i)\right\}
\end{equation}
i.e., only the intended receiver has sent an RTS and no other neighboring node 
has sent a CTS (except node $n_i$ itself). Finally, if $n_i$ is neither a PT 
nor an RR, it does not transmit.

As a final consideration, it is worth highlighting that full--duplex wireless 
and frequency--based channel access are strongly related in this protocol. On 
the one hand, FD capabilities are crucial to achieve an accurate simultaneous 
transmission and reception on different SCs, thereby enabling the proposed 
contention 
mechanism. On the other hand, this mechanism enables secondary transmissions 
concurrent to primary ones, thus allowing to fully exploit FD capabilities.
 
\subsection{Examples of operation}

\begin{figure}[b!]
\begin{center}
\scalebox{0.8}{
\begin{tikzpicture}
	\draw[fill=red!50!white,draw=black] (0,-1) circle (8pt);
	\draw[fill=red!50!white,draw=black] (3,-1) circle (8pt);
	\draw[fill=red!50!white,draw=black] (1.5,0.5) circle (8pt);
	\draw[dashed,very thick] (0.25,-.85) -- 
	(1.35,0.25);
	\draw[-latex,draw=red!50!white,fill=red!50!white,very thick] (0.15,-.75) -- 
	(1.25,0.35);
	\draw[dashed,very thick] (1.65,0.25) -- (2.75,-.85);
	\draw[-latex,draw=red!50!white,fill=red!50!white,very thick] (2.85,-.75) -- 
		(1.75,0.35);
	\node at (0,-1) {$n_1$};
	\node at (1.5,0.5) {$n_2$};
	\node at (3,-1) {$n_3$};
	\node at (1.5,-1.75) {(a) Scenario 1};
	\draw[fill=red!50!white,draw=black] (6,-1) circle (8pt);
	\draw[fill=red!50!white,draw=black] (9,-1) circle (8pt);
	\draw[fill=red!50!white,draw=black] (7.5,0.5) circle (8pt);
	\draw[dashed,very thick] (6.25,-.85) -- 
	(7.35,0.25);
	\draw[-latex,draw=red!50!white,fill=red!50!white,very thick] (6.15,-.75) -- 
	(7.25,0.35);
	\draw[latex-,draw=red!50!white,fill=red!50!white,very thick] (6.3,-1) -- 
	(7.5,0.2);
	\draw[dashed,very thick] (7.65,0.25) -- (8.75,-.85);
	\node at (6,-1) {$n_1$};
	\node at (7.5,0.5) {$n_2$};
	\node at (9,-1) {$n_3$};
	\node at (7.5,-1.75) {(b) Scenario 2};
	\draw [dashed, very thick] (3.25,0.5) -- (4,0.5);
	\draw [-latex, draw=red!50!white,fill=red!50!white,very thick] (3.25,0) -- 
	(4,0);
	\node at (4.75,0.5) [align=left] {TX range};
	\node at (5,0) [align=left] {TX intention};
\end{tikzpicture}}
\end{center}
\caption{Topology and transmission intentions for the operation examples.}
\label{fig:setup_example}
\end{figure}
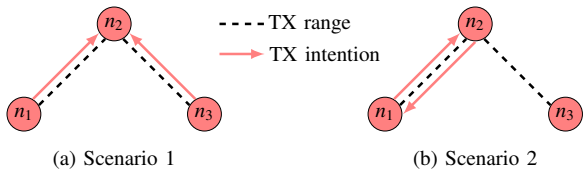

For a better understanding of how the proposed MAC strategy works, we provide 
here a couple of examples, for a simplified system with $N=3$ nodes and 
$S=6$ OFDM subcarriers. The simplest scheme for SC mapping is adopted, i.e.  
$\Sa=\{s_1,s_2,s_3\}$, $\Sb=\{s_4,s_5,s_6\}$, 
$\Fa(n_i)=s_i,\;\Fb(n_i)=s_{i+3},\;i=1,2,3$. 

Fig.~\ref{fig:sc_scenario1} and Fig.~\ref{fig:sc_scenario2} show the contention 
rounds for two different scenarios, respectively, while 
Fig.~\ref{fig:setup_example} reports the network topology and the transmission 
intentions.
In both scenarios, node $n_2$ is within the transmission range of nodes $n_1$ 
and $n_3$ that, however, cannot sense each other (two collision domains). In 
the first scenario, nodes $n_1$ and $n_3$ both intend to send a packet to 
$n_2$, resembling a typical hidden terminal situation. In the second one, nodes 
$n_1$ and $n_2$ have a packet for each other, opening the way for FD 
communications.

As reported in Fig.~\ref{fig:sc_scenario1} for Scenario 1, in the first round 
the two nodes randomly select two SCs and suppose that $\bar{s}_1=s_4$ and 
$\bar{s}_3=s_5$, with the result that both $n_1$ and $n_3$ are selected as PTs, 
since they can not sense each other's transmissions. 
Consequently, in the second round they both transmit, causing $n_2$ to hear 
signals on SCs $s_1$, $s_3$ and $s_5$. According to \eqref{eq:rr}, $n_2$ is 
selected as RR and transmits, during the third round, on SCs $s_2$ and $s_4$. 
Finally, according to \eqref{eq:transm_pt}, node $n_1$ is allowed to transmit, 
whereas the transmission by node $n_3$ is forbidden, since 
$\SC_{3,2}^3=\{s_4\}$ and $\Fb(n_3)=s_6$. It can be observed that the HT 
problem has been identified and solved thanks to the RCFD strategy.

In Scenario 2, as depicted in Fig.~\ref{fig:sc_scenario2}, nodes $n_1$ and 
$n_2$ participate in the first contention round, randomly selecting 
$\bar{s}_1=s_3$ and $\bar{s}_2=s_5$, therefore only $n_1$ is selected as PT. In 
the second round, $n_1$ transmits on SCs $s_1$ and $s_5$, thus node $n_2$ is 
selected as RR. Finally, in the third round $n_2$ transmits on SCs $s_2$ and 
$s_4$, providing a CTS to node $n_1$. 
Since the conditions in \eqref{eq:transm_pt} are verified for $n_1$ and those 
in \eqref{eq:transm_rr} are fulfilled for $n_2$, both nodes are cleared to 
transmit, thus enabling full--duplex transmission. We note 
that if node $n_2$ had been selected as PT in the first round, the final 
outcome would have been the same ($n_1$ selected as RR and subsequently cleared 
to transmit).

\subsection{Points of Discussion and Enhancements}
\label{subsec:discussion}

We recall from Section~\ref{subsec:assumptions} that the subcarrier mapping  
imposes a constraint on the number of nodes in the network, which has to be 
less than or equal to $S/2$.
However, it is worth stressing that the trend in wireless networks based on 
the IEEE 802.11 standard is to use wider channels, that offer an ever
increasing number of SCs. As an example, IEEE 802.11ac introduces 
80 MHz channels, that can accomodate 256 SCs, and allow up to 128 nodes.
Moreover, given a certain number of subcarriers $S$, we can do better than 
allowing only $S/2$ users if we plan to exploit the information carried in each 
SC. Indeed, in the presented algorithm only the presence or absence of data
on an SC was taken into account, whereas a more refined version can be 
implemented by taking the transmitted symbols in each SC into consideration. 
With this in mind, a more complex mapping between nodes and SCs can be 
performed that allows up to $m\cdot\frac{S}{2}$ users, since each SC can carry 
$\log_2 m$ bits if an $m$-ary modulation is used. As an example, if $S=64$ SCs 
are available and a BPSK modulation is employed, the system can host up to 64 
users. 

Another issue with the implementation of the RCFD protocol arises 
when asynchronous channel access is considered. As a matter of fact, in real 
networks nodes often generate packets and therefore try to access the channel 
in an independent manner. As a consequence, with the proposed algorithm 
implemented 
in a network with multiple collision domains, it could happen that a node 
starts a contention procedure while another node within its coverage range is 
receiving data, thus causing a collision. Indeed, the scanning procedure 
performed before the contention rounds is only capable of determining if a node 
within that range is transmitting, not if it is receiving.

To cope with this issue, we can make a simple yet effective modification to the 
algorithm, so that an idle node (which does not have a packet to send) that has already heard a 
CTS, refrains from accessing the channel until 
the end of the transmission is advertised through an ACK packet. To prevent 
freezing (in case the ACK is lost), a timeout can be started upon CTS detection 
and the node can again access the channel after its expiration.

\section{Performance Evaluation}
\label{sec:simulation}

\begin{figure*}[!ht]
	\centering
	\scalebox{0.9}{
	\subfloat[Normalized system 	
	throughput]{\includegraphics[width=\columnwidth]{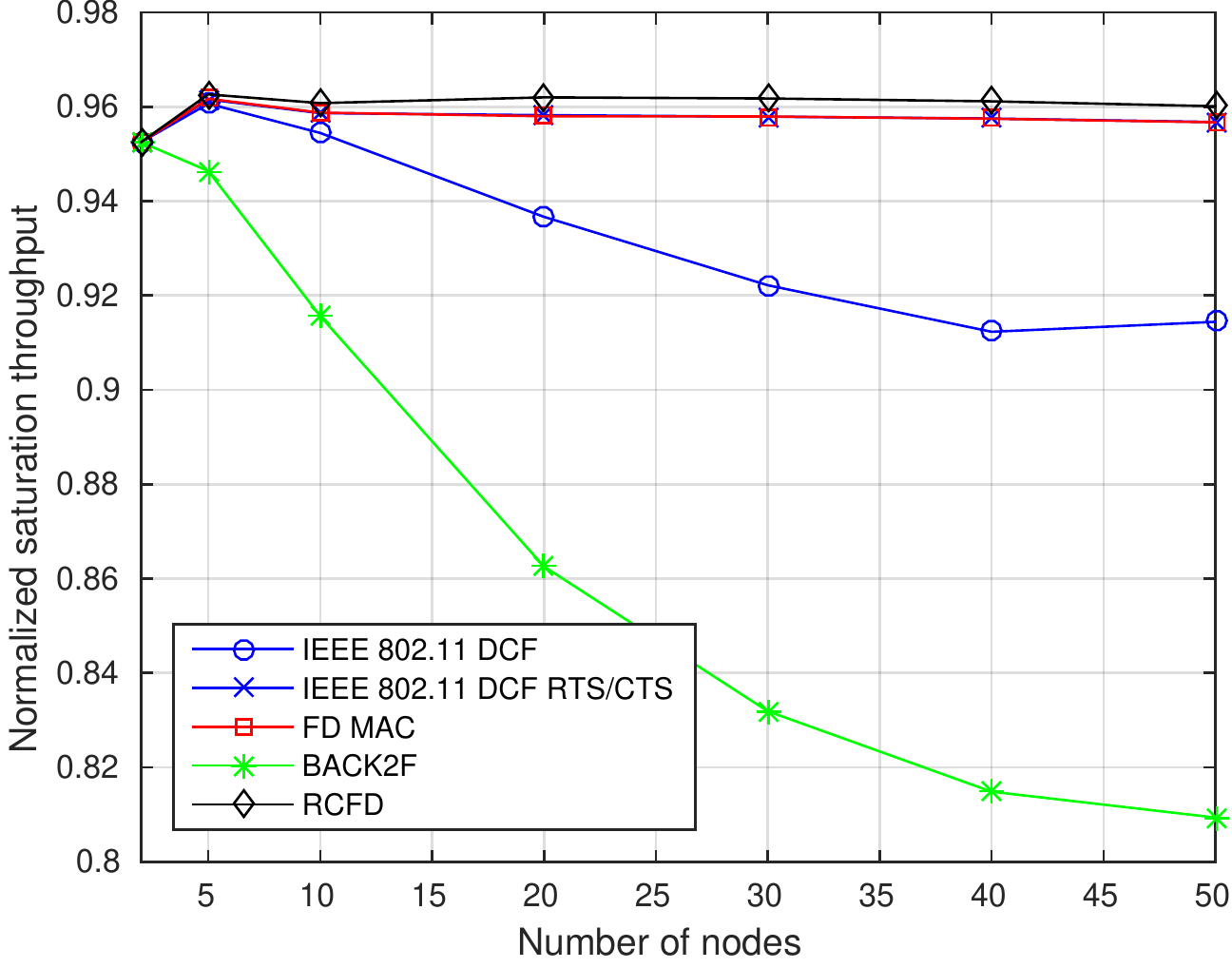}\label{fig:simulation_throughput_HL}}
	\hspace{1.2cm}
	\subfloat[Average 
	delay]{\includegraphics[width=\columnwidth]{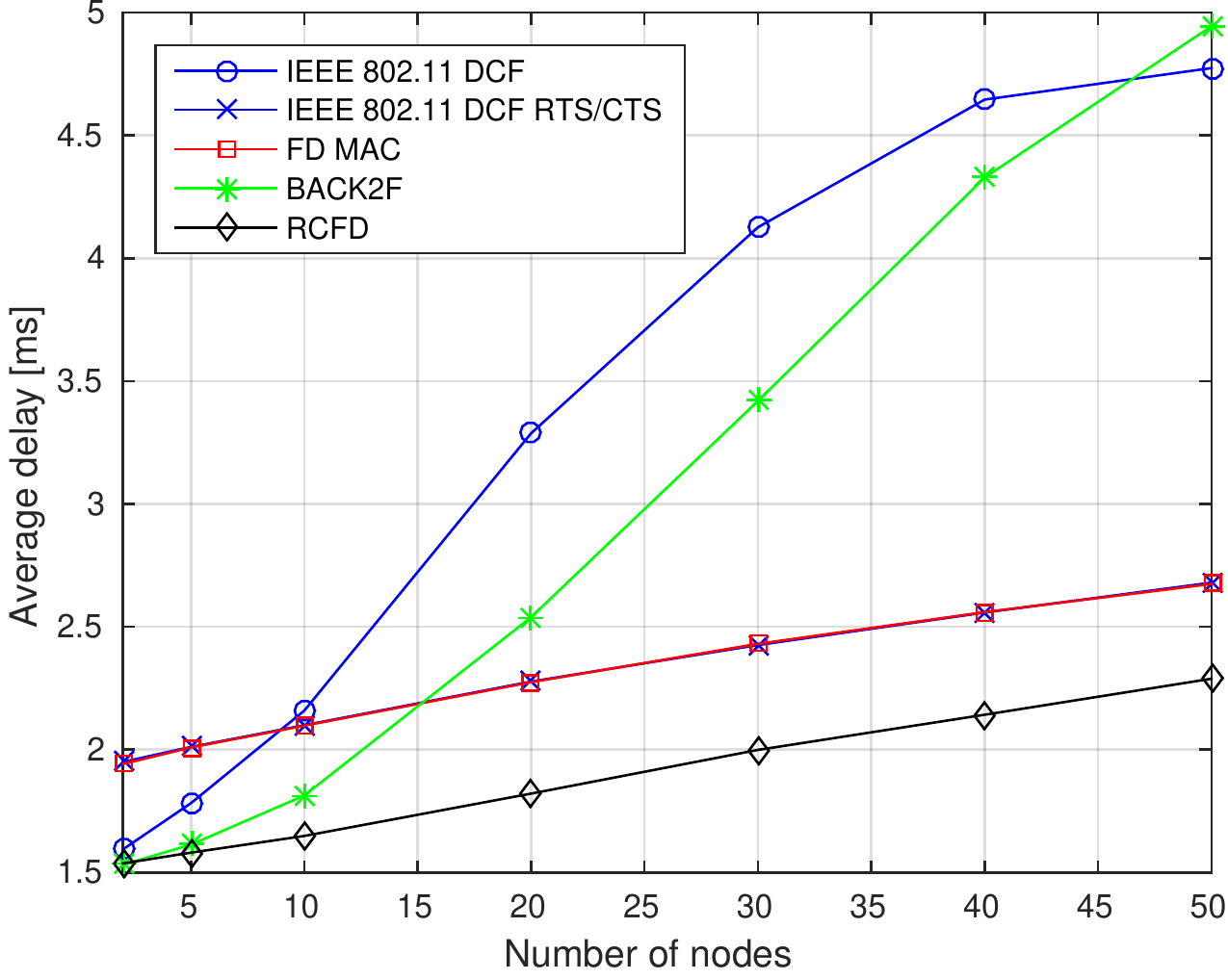}\label{fig:simulation_delay_HL}}}
	\vspace{1mm}
	\caption{Results of the Matlab simulations with long data transmission time 
	($R_T=1$~Mbit/s, $L=1000$~bits). The RCFD and FD MAC strategies always 
	offer a high system throughput regardless of the network size and the 
	former achieves the lowest average delay.}
	\label{fig:simulation_HL}
\end{figure*} 

The proposed RCFD protocol is evaluated and compared to other, standard and 
state--of--the--art, MAC layer protocols through a network simulator developed 
in Matlab,\footnote{Further simulations based on \textit{ns3} are available in 
an extended version of this work \cite{rcfd_journal}.} able to mimic the 
exchange of packets in a full--duplex ad hoc 
wireless network. Specifically, the Distributed Coordination Function (DCF) 
presented in the IEEE 802.11 standard \cite{ieee80211std} is regarded as the 
baseline MAC strategy, also considering its version with the RTS/CTS option 
enabled. Among the various MAC proposals for full--duplex wireless networks 
discussed in Section~\ref{subsec:related_fd}, we selected the FD MAC strategy of
 \cite{Duarte2014} since it is one of the most general approaches and does not 
require assumptions on network topology, traffic pattern or PHY configuration. 
Finally, the BACK2F scheme \cite{Sen2011} has been chosen as an example of 
protocols that perform channel contention in the frequency domain.

\subsection{Simulation Setup}
\label{subsec:simulation_setup}

Tab.~\ref{tab:sim_param} presents the values of all the parameters adopted in 
the simulations.

\begin{table}[!b]
\caption{Simulations parameters}
\label{tab:sim_param}
\centering
\scalebox{0.9}{
\begin{tabular}{l|c|c}
\toprule
{\bfseries Parameter} & {\bfseries Description} & {\bfseries Value}\\
\midrule
$N$ & Number of nodes in the network & $\{2,5,10,20,30,40,50\}$\\
$R_S$ & Source rate of each node & 10~Kbit/s\\
$L$ & Average payload size & $\{200,1000\}$ bits\\
$R_T$ & PHY layer transmission rate & $\{1,54\}$ Mbit/s\\
$S$ & Number of OFDM subcarriers & $64$\\
$M$ & Number of different network realizations & 100\\
$T$ & Duration of each simulation & 10~s\\
\bottomrule
\end{tabular}}
\end{table}

We simulated a network of $N$ nodes randomly deployed over a unit square 
according to a uniform distribution. Each node has a coverage radius of 
$r(N)=\sqrt{\frac{2}{N}\log N}$ \footnote{This value is chosen in order 
to guarantee that the network is connected with high probability 
\cite{Gupta2000}.}, and can transmit only to, and overhear any transmission by, 
nodes within this range. 

Packets are generated at each node according to a Poisson process, whose 
arrival rate is computed based on two simulation parameters, namely the source 
rate $R_S$ (expressed in bit/s) and the average payload size in bits $L$.
All nodes have FD capabilities, adopt an IEEE 802.11g PHY layer and transmit 
the data with the same PHY transmission rate $R_T$.

For the frequency--domain based algorithms, we have considered $S$ OFDM 
subcarriers. In the specific case of the RCFD algorithm, we implemented the 
more sophisticated mapping mentioned in Section~\ref{subsec:discussion} to 
allow a maximum number of users equal to $S$ (adopting BPSK). We 
have also implemented the deferring scheme discussed in the same Section, since 
in the simulations nodes access the channel asynchronously.

For each choice of the number of nodes $N$, a total of $M$ simulations were 
performed, each with a different node distribution and a specific realization 
of the packet generation processes. The operations of the five MAC algorithms 
have been compared over the same set of packets and node distribution for a 
total time $T=10$~s, during which hundreds of thousands of packets were 
generated. 
Results have then been averaged over all simulations. 

As to the performance metrics, we have first considered the \textit{system 
throughput} $G$, expressed as the ratio between the sum of the payload lengths 
of all packets successfully delivered and the time needed for the 
generation and transmission of all the packets. Specifically, we considered 
the \textit{normalized throughput}, expressed as
\begin{equation}
G_0 = \frac{G}{N\cdot R_S}
\end{equation} 
which represents the fraction of the generated traffic which is actually 
delivered.

The \textit{average delay} has also been taken into account, defined as the 
average time elapsed between the generation of a packet and the instant in 
which it is successfully delivered or permanently discarded. 

\subsection{Simulation Results}
\label{subsec:simulation_results}

\begin{figure*}[!t]
	\centering
	\scalebox{0.9}{
	\subfloat[Normalized system 	
	throughput]{\includegraphics[width=\columnwidth]{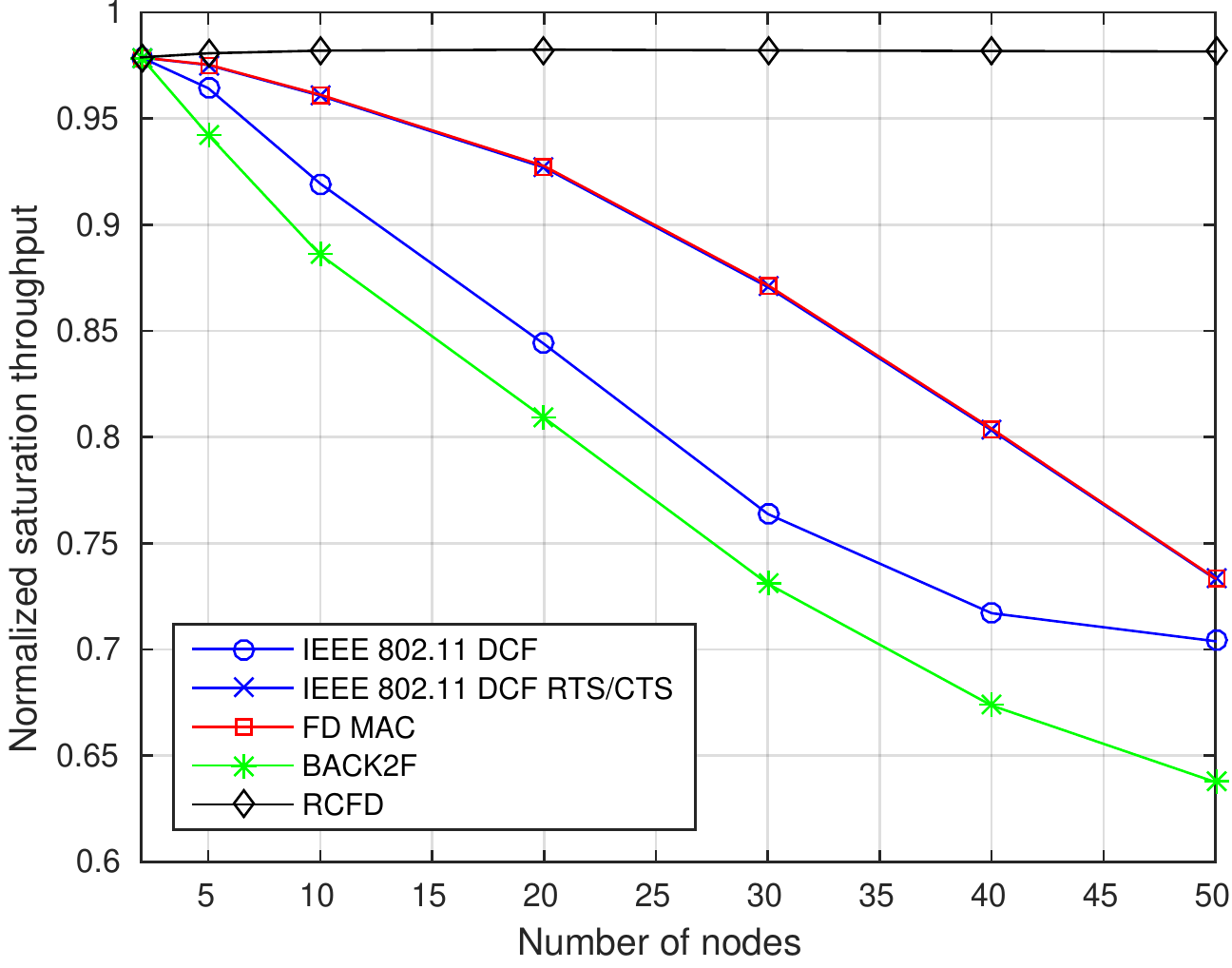}\label{fig:simulation_throughput_LL}}
	\hspace{1.2cm}
	\subfloat[Average 	
	delay]{\includegraphics[width=\columnwidth]{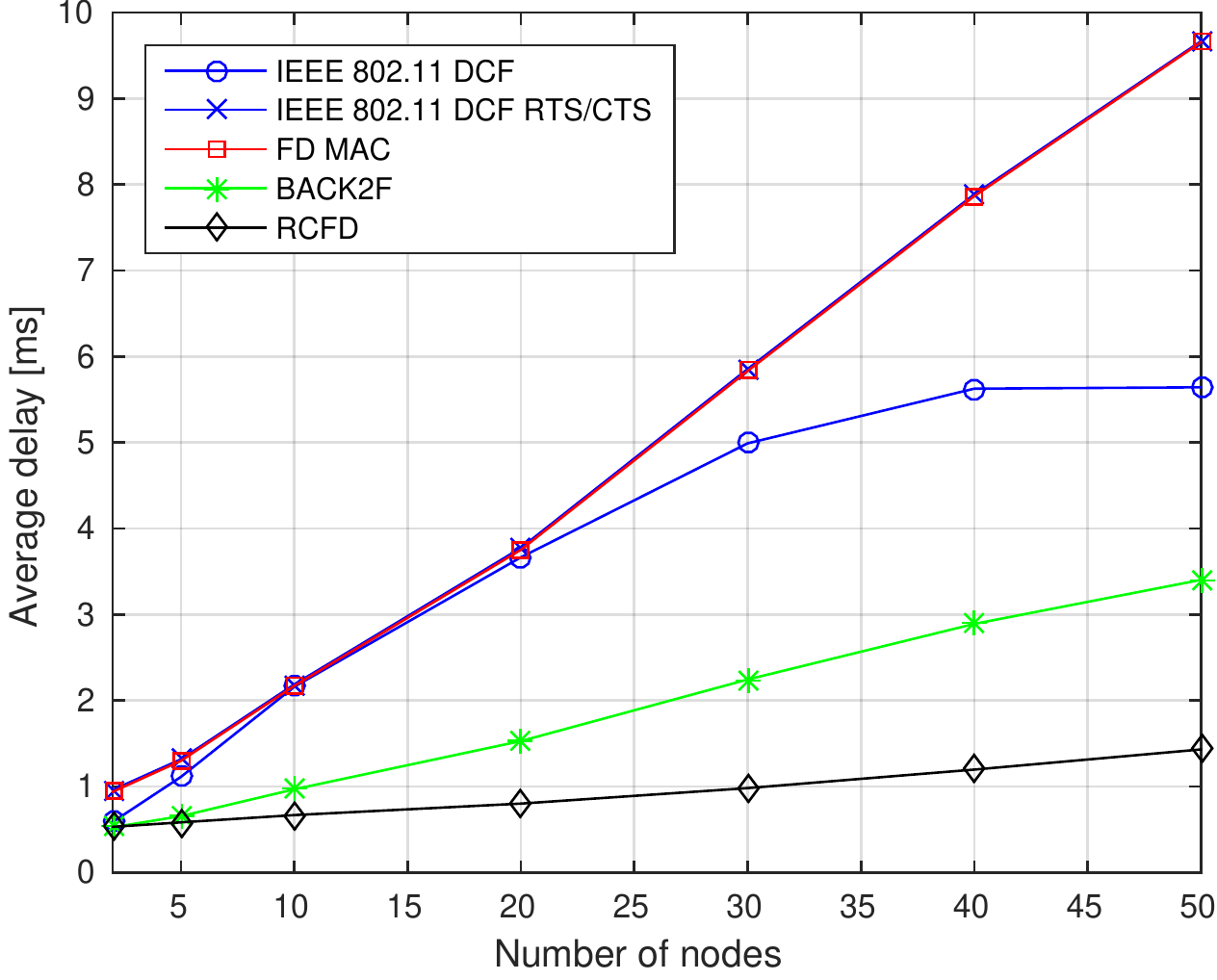}\label{fig:simulation_delay_LL}}
	}
	\vspace{1mm}
	\caption{Results of the Matlab simulations with short data transmission 
	time ($R_T=54$~Mbit/s, $L=200$~bits). The RCFD strategy is the 
	only one that always offers a high system throughput regardless of the 
	network size and achieves the lowest average delay.}
	\label{fig:simulation_LL}
\end{figure*} 

We have evaluated two different scenarios in our network simulations. In each 
scenario, the aforementioned performance metrics for the considered MAC 
algorithms have been evaluated for a different number of nodes in the network. 
The scenarios differ for the average duration of the data transmission, 
expressed as the ratio between the average payload size $L$ and the data 
transmission rate $R_T$. 

The results are reported in terms of normalized system throughput and average 
delay in Fig.~\ref{fig:simulation_HL} for the case of long average data 
transmission time, i.e., $L=1000$~bits and $R_T=1$~Mbit/s, yielding an 
average data transmission time of 1~ms.  
It can be noticed that the RCFD algorithm provides the best performance 
in terms of both throughput and delay. 
As far as throughput is concerned, the performance of RCFD does not degrade 
when 
the number of nodes in the network increases. The same behavior can be observed 
for the two algorithms based on time--domain RTS/CTS (FD MAC and IEEE 802.11), 
which provide a normalized throughput almost equal to that of RCFD. 
Conversely, both standard IEEE 802.11~DCF and BACK2F decrease their performance
as the network becomes denser, with the latter suffering nearly a 15$\%$ 
throughput degradation for a large network size. 
A similar situation can be observed when taking average delay into account. 
Indeed, RCFD yields the lowest delay, outperforming the two strategies based on 
time--domain RTS/CTS by roughly 33$\%$ as for all of them the delay increases 
linearly with the number of nodes. On the contrary, the average delay in IEEE 
802.11 and BACK2F increases much more significantly with the network size, 
especially for the case of large network size, where the average delay is more 
than doubled with respect to RCFD.

The outcomes for the other scenario are reported in 
Fig.~\ref{fig:simulation_LL}, for an average payload length of 
$L=200$~bits and a raw data rate of $R_T=54$~Mbit/s, thus yielding a packet 
transmission time of 3.7~\mus.
With respect to the previous scenario, in this case the strategies based on 
time--domain RTS/CTS (FD MAC and IEEE 802.11) perform significantly worse, 
since the impact of the overhead caused by additional frame exchanges is much 
more significant. Indeed, looking at the throughput in 
Fig.~\ref{fig:simulation_throughput_LL}, it can be observed that in these two 
strategies the performance significantly degrades when the network size 
increases, with a decrease of more than 20$\%$ for a large number of nodes. As 
far as the average delay is concerned, the performance degradation is even more 
evident, as the two time--domain strategies offer the worst performance among 
all the considered MAC algorithms. 
Conversely, our proposed RCFD algorithm is able to guarantee an almost constant 
normalized throughput (close to 1) regardless of the network size, and to 
ensure a very low average delay, always smaller than 2~ms. This is by far the 
best performance compared to all the other MAC layer algorithms considered.

\section{Conclusions}
\label{sec:conclusions}

In this paper we proposed RCFD, a full--duplex MAC protocol based on a
frequency--domain channel 
access procedure. We showed through simulation that this 
strategy provides excellent performance in terms of both 
throughput and delay, also in the case of dense networks with a large number of 
nodes, compared to other standard and state--of--the--art MAC layer schemes.

A natural, though challenging, extension of this work is the experimental 
validation of the proposed MAC layer protocol on devices capable of FD 
operations and able to transmit OFDM symbols using only some specific 
subcarriers. 
A theoretical analysis of the proposed scheme can also be envisioned, following
the analytical approaches that have been used to study traditional 
MAC protocols for wireless networks. 
Finally, the impact of a non--ideal channel must be investigated, especially 
for the case of impairments that may occur during the channel contention phase.
\vspace{-.2cm}

\bibliographystyle{IEEEtran}
\bibliography{biblio_fdw}
\end{document}